\documentclass[journal]{IEEEtran}

\ifCLASSINFOpdf
\else
   \usepackage[dvips]{graphicx}
\fi
\usepackage{url}

\usepackage{graphicx}
\usepackage{amsmath,amssymb,amsfonts}
\usepackage{amsthm}
\usepackage{comment}
\usepackage{amsbsy}
\usepackage{mathtools}
\usepackage{algorithm}
\usepackage{algpseudocode}
\usepackage{subcaption}
\usepackage{enumitem}
\begin{document}

\title{Outlier-resilient model fitting via percentile losses: Methods for general and convex residuals}

	\author{Jo\~ao Domingos and Jo\~ao Xavier
	\thanks{\hspace{-.61cm} Jo\~ao Domingos (oliveira.domingos@tecnico.ulisboa.pt) and João Xavier (jxavier@isr.tecnico.ulisboa.pt) are with Instituto de Sistemas e Robótica of Instituto Superior Técnico, Portugal.
		This work has been supported by FCT through LARSyS funding (DOI: 10.54499/LA/P/0083/2020, 10.54499/UIDP/50009/2020, and 10.54499/UIDB/50009/2020) and research grant with the reference PD/BD/150631/2020.}
}


\newtheorem{theorem}{Theorem}
\newtheorem{lemma}{Lemma}
\newtheorem{remark}{Remark}
\newtheorem{example}{Example}
\newtheorem{result}{Result}
\newtheorem{definition}{Definition}
\newtheorem{corollary}{Corollary}[theorem]
\maketitle

\begin{abstract}
We consider the problem of robustly fitting a model to data that includes outliers by formulating a percentile optimization problem. This problem is non-smooth and non-convex, hence hard to solve. We derive properties that the minimizers of such problems must satisfy. 

These properties lead to methods that solve the percentile formulation both for general residuals and for convex residuals. The methods fit the model to subsets of the data, and then extract the solution of the percentile formulation from these partial fits. 
As illustrative simulations show, such methods endure higher outlier percentages, when compared with standard robust estimates. 

Additionally, the derived properties provide a broader and  alternative theoretical validation for existing robust methods, whose validity was previously limited to specific forms of the residuals.
\end{abstract}


\IEEEpeerreviewmaketitle

\section{Introduction}
\label{sec:introduction}
\IEEEPARstart{W}{e} consider the problem of fitting a model to $M$ data points, where $O$ of them are outliers ($1 \leq O < M$). A data point is denoted by $(x_m , y_m)$. The vector $x_m \in \mathbf{R}^p$ is the feature and the scalar $y_m 
\in \mathbf{R}$ is the label, for  $m \in \mathcal{M} = \{ 1, 2, \ldots,  M \}$. The  model $h$ is parameterized by $\theta \in \mathbf{R}^d$ and predicts the label $y_m$ from the feature $x_m$.

Let $f_m(\theta)$ be the residual of the data point $(x_m, y_m)$ when the model parameter is set to $\theta$ (a common choice is  $f_m(\theta) = | y_m - h( x_m, \theta) |$). We thus face the problem of computing a model parameter $\theta$ that satisfies \begin{align}
 f_m(\theta) \simeq 0,
	\label{eqn:high_level_model}
\end{align}
for $M-O$ data points, but not necessarily for \emph{all} data points because of the outliers.  In other words,  letting $\mathcal{I} \subset \mathcal{M}$ index the data points that are inliers and $\mathcal{O} = \mathcal{M} \backslash \mathcal{I}$ the outliers (thus, $O = | \mathcal{O} |$), 
we seek a $\theta$ that satisfies \eqref{eqn:high_level_model} for $m \in \mathcal{I}$, but not necessarily for $m \in \mathcal{O}$. The challenge lies in that neither $\mathcal{I}$ nor $\mathcal{O}$ is given beforehand.  

\medskip

\noindent \textbf{Non-robust least-squares formulation.} 
If the presence of outliers is ignored, model fitting can be cast as the problem of minimizing a global quadratic loss of the residuals, 
\begin{align}
	\min_\theta \phi_{\text{quad}}\left( f_m(\theta)  ;\, m \in \mathcal{M} \right),
	\label{eqn:LS_formulation}
\end{align}
where $\phi_{\text{quad}}( z )$ gives the sum-of-squares of the components of its input vector $z = ( z_m \, ; \, 1 \leq  m \leq M  )$, 
\begin{align}
	\phi_{\text{quad}}(z) = \sum_{m = 1}^M  z_m^2,
	\label{eqn:LS_loss}
\end{align}
see~\cite{watson1967linear,van199310, dykstra1983algorithm,johnson19921,sorenson1970least}. 
Such methods, however, are not robust.
As soon as even just a few outliers are introduced, these methods lead to a $\theta$ that satisfies~\eqref{eqn:high_level_model} only for a very small number of data points~\cite{rousseeuw1984least}. 

\section{Robust Percentile Formulation}
\label{sec:formulation}

Finding a parameter $\theta$ that minimizes the residuals of inliers, while ignoring a number $O$ of outliers, can be captured by the optimization problem
\begin{align}
	\min_{\theta}  \phi_{\text{per}} \left( f_m(\theta) ; \, m \in \mathcal{M}\right).
	\label{eqn:percentile_min}
\end{align}
Here, $\phi_{\text{per}}$ is the percentile function of order $O$: The function $\phi_{\text{per}}$ 
discards the top $O$ components of the input vector $z$ and returns the largest component that remains. 
In other words, let $z = ( z_m \, ; \, 1 \leq m \leq M )$ be the vector inputted  into $\phi_{\text{per}}$ and let $\{ m_1, m_2, \ldots, m_{M} \}$ be a permutation of $\{ 1, 2, \ldots, M \}$ that sorts $z$ in descending order, 
\begin{align}
	z_{m_1}\geq \dots \geq z_{m_{O}} \geq z_{m_{O+1}} \geq \dots \geq z_{m_{M}}.
	\label{eqn:sort_elements}
\end{align}
We have $\phi_{\text{per}}(z) = z_{m_{O+1}}$. Note that the permutation $\{ m_1, m_2, \ldots, m_{M} \}$ depends on $z$: the function $\phi_{\text{per}}$ is non-linear.

In~\eqref{eqn:percentile_min}, the effectiveness of a candidate parameter~$\theta$ is thus gauged by its ability to best fit $M-O$ data points, since its $O$ poorest fits are excluded.


The percentile formulation~\eqref{eqn:percentile_min} is known in different communities by different names. We mention two examples:
\paragraph{Least Quantile of Squares (LQS) regression}
When each residual is taken as $f_m(\theta) = | y_m - h( x_m, \theta) |$, with the model being linear ($h(x_m, \theta) = x_m^T\,\theta$), and 
the dataset contains about $50\%$ of outliers, $O \simeq M/2$, the percentile loss $\phi_{\text{per}}$ becomes the median and~\eqref{eqn:percentile_min} collapses into the least median of squares (LMS) formulation in~\cite{rousseeuw1984least}. The LMS formulation has been  generalized to arbitrary percentiles beyond $50\%$, being known as Least Quantile of Squares (LQS) regression~\cite{rousseeuw1997recent}.  
Although solving the LMS formulation  is NP-HARD~\cite{bernholt2006robust}, efficient methods have been proposed for the case of intersect-slope regression~\cite{rousseeuw1984least,souvaine1987time,shapira2015gpu,mount2007practical,edelsbrunner1990computing}. Exact methods for small parameter size $d$ are given in~\cite{stromberg1991computing,agullo1997exact,giloni2002least,Bertsimas_modern}.   Section~\ref{sec:stromberg} connects our results
with Stromberg's~method~\cite{stromberg1991computing} for the LQS problem.


\paragraph{Value-at-Risk (VaR) optimization}
Formulation~\eqref{eqn:percentile_min} also has a strong connection to portfolio optimization~\cite{Quantil_Regression_2001,Regression_Quantiles_1978,CVAR_1999}, since~\eqref{eqn:percentile_min} can be interpreted as optimizing the Value-at-Risk (VaR) measure of an underlying stochastic risk problem. Further details on this interpretation for the setting of target localization can be found in~\cite{domingos2023robust}.

\section{Contributions}
Solving~\eqref{eqn:percentile_min} is generally NP-HARD~\cite{bernholt2006robust}, if only because the percentile loss $\phi_{\text{per}}$ is by itself already non-smooth and non-convex. This section derives properties for the minimizers of~\eqref{eqn:percentile_min} for two cases: for general residuals, in the sense that each $f_m(\theta)$ is not necessarily convex in $\theta$; and for convex residuals, where each $f_m(\theta)$ is convex in $\theta$. These properties are relevant because they reveal principled ways of solving some instances of~\eqref{eqn:percentile_min} via a finite (yet exponential) number of convex programs. Although the exponential dependency can often be relaxed by randomization~\cite{rousseeuw1997recent,chakraborty2008optimization,joss1990probabilistic}, this letter is mainly devoted to solving~\eqref{eqn:percentile_min} exactly. 

\medskip

\noindent \textbf{General residuals.} We show that~\eqref{eqn:percentile_min} can be solved by solving 
\begin{align}
\label{eqn:pfit}
	\min_\theta \phi_{\text{max}}\left( f_m(\theta)  ;\, m \in \mathcal{S} \right),
\end{align}
for all subsets $\mathcal{S}$ of $\mathcal{M}$ whose cardinality $S = | \mathcal{S} |$ is $M-O$ (the number of inliers), where $\phi_{\max}$ is the worst-case loss: for a generic vector $z = ( z_s ; \, 1 \leq s \leq S)$, 
\begin{align}
	\phi_{\text{max}}(z) = \max\left\{ z_1, z_2, \ldots, z_S \right\}.
	\label{eqn:WC_loss}
\end{align}
We refer to~\eqref{eqn:pfit} as an ${\mathcal S}$-fit. In such a fit, the model is adjusted only to the $S$ data points indexed by $\mathcal{S}$, not to the full data set~$\mathcal{M}$. 

We show that a $\theta$ that yields the best $\mathcal{S}$-fit is also a solution of~\eqref{eqn:percentile_min}---see Theorem 1.

The strategy of seeking the best $\mathcal{S}$-fit of size $M-O$ is not new in robust estimation -- see ~\cite{agullo1997exact,cook1993exact,van2009minimum,Bertsimas_modern,stromberg1993computing}. However this idea is often explored in dedicated setups with no connection across problems. From our perspective, the contribution of Theorem 1 is essentially two-fold:
\begin{itemize}
	\item To the best of our knowledge, this letter presents the first proof that using ${\mathcal S}$-fits is always optimal, regardless of the residuals $f_m$. Bertsimas et. al~\cite{Bertsimas_modern} (Theorem 4.1) proved this result for LQS regression. Their proof explores a mixed integer formulation of~\eqref{eqn:percentile_min} for the residual form  $f_m(\theta) = | y_m - h( x_m, \theta) |$. Theorem 1 is also mentioned (without proof) in~\cite{agullo1997exact,stromberg1993computing} for LQS regression and in~\cite{cook1993exact,van2009minimum}  for minimum volume ellipsoids. So Theorem 1 unifies several results in robust estimation by giving a formal proof that ${\mathcal S}$-fits are always optimal.
	\item As a secondary contribution, Theorem 1 allows us to interpret several algorithms~\cite{hawkins1993feasible,agullo1997exact,watson1998computing} that approach problem~\eqref{eqn:percentile_min} by alternating between minimax fits and updates of set $\mathcal{S}$. Equality~\eqref{eq:id1} of Theorem 1 reveals the main motivation for these updates.
\end{itemize}
\medskip

\noindent \textbf{Convex residuals.} Assuming convex residuals and $d + 1 < M-O$, we show that the solution of~\eqref{eqn:percentile_min} must also be a solution of some $\mathcal{S}$-fit~\eqref{eqn:pfit},  but where $\mathcal{S}$ has now a smaller size, namely,  $d+1$. This means that, once the solutions of all such $\mathcal{S}$-fits are available, it suffices to evaluate them in the objective~\eqref{eqn:percentile_min} and retain the best to arrive  at a solution of~\eqref{eqn:percentile_min}---see Theorem~2.

This strategy of looking for the solution of~\eqref{eqn:percentile_min} among the solutions of $\mathcal{S}$-fits of size $d+1$ is less understood in robust statistics~\cite{stromberg1993computing,van2009minimum}, in the sense that Theorem 2 was only proved for LQS regression with $f_m(\theta) = | y_m - x_m^T \theta |$ -- see~\cite{stromberg1993computing} for the original proof. From our perspective, Theorem 2 is the main contribution of the paper since sampling  $\mathcal{S}$ sets of size $d+1$ can be tractable in low dimensional setups -- see Section~\ref{sec:stromberg}.

\section{General residuals}
\label{sec:gen_res}

For a generic function $\psi\,\colon\,\mathbf{R}^n \rightarrow \mathbf{R}$, its infimum is denoted by $\psi^\star$ and its set of global minimizers by $\text{Argmin}(\psi)$. That is, $\psi^\star = \textbf{inf} \left\{ \psi(x) \, \colon \, x \in \mathbf{R}^n \right\}$ and $\text{Argmin}(\psi) = \left\{ x \in \mathbf{R}^n \, \colon \, \psi(x) = \psi^\star \right\}$.

Hereafter, we let $\phi$ denote the objective function in~\eqref{eqn:percentile_min} and $\phi_{\mathcal{S}}$  the objective function in~\eqref{eqn:pfit}.

\begin{theorem}
	\label{theorem:subset_sampling}
We have \begin{align}
\phi^\star = \min_{| \mathcal{S} | = M-O}\,  \phi_\mathcal{S}^\star   
\label{eq:id1}
\end{align}
and
\begin{align}
    \text{Argmin}({\phi}) = \text{Argmin}({\phi_{\mathcal{S}^\star}}),
    \label{eq:id2}
\end{align}
where $\mathcal{S}^\star$ is any $\mathcal{S}$ that achieves the minimum on the right-hand side of~\eqref{eq:id1}.
\end{theorem}

~

Property~\eqref{eq:id2} says that the solutions of the percentile formulation~\eqref{eqn:percentile_min} are exactly the solutions of the best $\mathcal{S}$-fit. This gives a way to solve~\eqref{eqn:percentile_min}: first, solve all $\mathcal{S}$-fits~\eqref{eqn:pfit}, thus accessing $\text{Argmin}( \phi_{\mathcal{S}} )$ and $\phi_{\mathcal{S}}^\star$ for all $\mathcal{S}$ of size $M-O$;  then,  locate the best $\mathcal{S}$, say, $\mathcal{S}^\star$, in the sense that $\phi_{\mathcal{S}^\star }^\star \leq  \phi_{\mathcal{S}}^\star$ for all $\mathcal{S}$; finally, pick $\theta^\star \in \text{Argmin}( \phi_{\mathcal{S}^\star} )$. Such a~$\theta^\star$ solves the percentile formulation.


\medskip

\noindent \textbf{Proof of Theorem  1.} 
We make use of following fact, whose straightforward proof is ommited: if $\psi \, \colon \, \mathbf{R}^n \rightarrow \mathbf{R}$ is a function of the form $$\psi(x) = \min\left\{\psi_1(x), \psi_2(x), \ldots, \psi_K(x) \right\}, \quad \text{for all }x,$$ where $\psi_k \, \colon \, \mathbf{R}^n \rightarrow \mathbf{R}$, then 
\begin{align}
    \psi^\star = \min\left\{ {\psi}_1^\star, {\psi}_2^\star, \ldots, \psi_K^\star \right\},
    \label{eq:infpsi}
\end{align}
and 
\begin{align}
    \text{Argmin}( \psi ) = \text{Argmin}( \psi_{k^\star} ),
    \label{eq:infpsi2}
\end{align}
where $k^\star$ is any $k$ that achieves the minimum on the right-hand side of~\eqref{eq:infpsi}, that is, $\psi_{k^\star}^\star \leq \psi_k^\star$ for $1 \leq k \leq K$.

We now turn to the proof of Theorem 1. We start by noting that just evaluating the percentile function $\phi_{\text{per}}$ at a given vector $z = ( z_m ; 1 \leq m \leq M )$ already reduces to seeking subsets of $\{ 1, 2, \ldots, M \}$ of size $M-O$ for the minimal largest component:
\begin{align}
	\phi_{\text{per}}(z)=\min_{|\mathcal S| = S}\, \phi_{\max}\left( z_m ; m \in \mathcal{S} \right),
	\label{eqn:percentile_reformulation}
\end{align}
where $S = M-O$.

To prove~\eqref{eqn:percentile_reformulation} consider the permutation $\{ m_1, m_2, \ldots, m_M \}$ of $\{ 1, 2,  \ldots, M \}$ that sorts the components of the given $z$ in descending order, 
\begin{align}
	z_{m_1} \geq \cdots \geq z_{m_{O}} \geq z_{m_{O+1}} \geq \cdots \geq z_{m_M}.
	\label{eqn:permutation_Def}
\end{align}
Recall that $\phi_{\text{per}}(z) = z_{m_{O+1}}$.

Now consider the particular choice of subset $\mathcal{S}^\star = \{ m_{O+1}, \ldots, m_M \}$, which implies $\phi_{\max}( z_m ; \, m \in \mathcal{S}^\star ) = z_{m_{O+1}}$. This shows that the inequality $\geq$ holds in~\eqref{eqn:percentile_reformulation}.

It remains to show that the reverse inequality $\leq$ also holds. For this, take an arbitrary subset $\mathcal{S} \neq \mathcal{S}^\star$ of size $S$, say, $\mathcal{S} = \{ j_1, j_2, \ldots, j_S \}$. Because $\mathcal{S} \neq \mathcal{S}^\star$ there exists an $1 \leq s \leq S$ such that $j_s \in \mathcal{S} \backslash \mathcal{S}^\star$, which then necessarily satisfies $z_{j_s} \geq z_{m_{O+1}}$ (if $z_{m_{O+1}} > z_{j_s}$ were to hold, then $j_s$ would have to be in $\{ m_{O+2}, \ldots, m_M \}$, a contradiction to $j_s \not\in \mathcal{S}^\star$). We thus have $\phi_{\max}( \mathcal{S} ) \geq z_{j_s} \geq z_{m_{O+1}}$. Since $\mathcal{S}$ was chosen arbitrarily, this shows that the inequality $\geq$ holds in~\eqref{eqn:percentile_reformulation}.

Using characterization~\eqref{eqn:percentile_reformulation}, we thus have
\begin{align}
    \phi_{\text{per}}( f_m(\theta) ; \, m \in \mathcal{M} ) = \min_{| \mathcal{S} | = S} \, \phi_{\max}( f_m(\theta) ; \, m \in \mathcal{S} ), \nonumber
\end{align}
for all $\theta$, or, more compactly,
\begin{align}
    \phi = \min_{| \mathcal{S} | = S} \phi_{\mathcal{S}}.
    \label{eq:key}
\end{align}
Identity~\eqref{eq:key}, together with~\eqref{eq:infpsi} and~\eqref{eq:infpsi2}, imply the claimed properties~\eqref{eq:id1} and~\eqref{eq:id2}. 

\section{Convex residuals}
\label{sec:cvx_res}

We now turn to convex residuals.

\begin{theorem}
	\label{theorem:cvx_res}
If $f_m(\theta)$ is convex in $\theta$ for all~$m$, and $d+1 < M-O$, then 
\begin{align}
\text{Argmin}( \phi ) \subset \bigcup_{| \mathcal{S} | = d+1}\, \text{Argmin}( \phi_{\mathcal{S}} ).
    \label{eq:id_cvx}
\end{align}
\end{theorem}

Property~\eqref{eq:id_cvx} asserts that the solutions of the percentile formulation~\eqref{eqn:percentile_min} are to be found among the solutions of the $\mathcal{S}$-fits~\eqref{eqn:pfit}, where now $\mathcal{S}$ has size $d+1$. This enables~\eqref{eqn:percentile_min} to be solved as follows: first, solve all $\mathcal{S}$-fits, which are now convex optimization problems, thus accessing $\text{Argmin}( \phi_{\mathcal{S}} )$ for all $\mathcal{S}$ of size $d+1$; then, plug all $\theta$ from the sets $\text{Argmin}( \phi_{\mathcal{S}} )$ in the objective function of interest $\phi$, and call $\theta^\star$ one that achieves the lowest value. Such $\theta^\star$ solves the percentile formulation. This strategy is most direct to carry out when each $\text{Argmin}( \phi_{\mathcal{S}} )$ is a singleton: see section VI for further discussion.

\medskip

\noindent \textbf{Proof of Theorem  2.} We start by noting that property~\eqref{eq:id2} from Theorem 1 implies 
\begin{align}
    \label{eq:step1}
    \text{Argmin}( \phi ) \subset \bigcup_{| \mathcal{T }| = M-O} \text{Argmin}( \phi_{\mathcal{T}} ).
\end{align}
Thus, we need only show that, for any given $\mathcal{T} \subset \mathcal{M}$ of size $M-O$, we can find a $\mathcal{S} \subset \mathcal{M}$ of size $d+1$ such that \begin{align}
    \text{Argmin}( \phi_{\mathcal{T}} ) \subset \text{Argmin}( \phi_{\mathcal{S}} ).
\end{align}

Let $\mathcal{T}$ of size $M-O$ be given and choose $\theta^\star \in \text{Argmin}( \phi_{\mathcal{T}} )$. Our goal is thus to come up with a set $\mathcal{S}$ of size $d+1$ such that \begin{align}
\theta^\star \in \text{Argmin}( \phi_{\mathcal{S}} ).
    \label{eq:goal}
\end{align}

Because $\theta^\star$ solves the convex optimization problem 
\begin{align}
	\min_\theta \phi_{\mathcal{T}}(\theta), \nonumber
\end{align}
Theorem 2.2.1 in~\cite{hiriart2004fundamentals} states that \begin{align}
    \label{eq:stat}
    0 \in \partial \phi_{\mathcal{T}}\left( \theta^\star \right),
\end{align}
where the set $\partial \phi_{\mathcal{T}}\left( \theta^\star \right)$ denotes the sub-differential of the convex function $\phi_{\mathcal{T}}$ at the point $\theta^\star$.

Because $\phi_{\mathcal{T}}$ is a pointwise maximum of convex functions, namely, $$\phi_{\mathcal{T}}(\theta) = \max\{ \phi_m(\theta) ; m \in \mathcal{T} \}, \quad \text{for all }\theta,$$
we can use Corollary 4.3.2 in~\cite{hiriart2004fundamentals}, together with~\eqref{eq:stat}, to conclude that
\begin{align}
    0 \in \text{co} \left( \bigcup_{m \in \mathcal{A}^\star} \partial \phi_m( \theta^\star ) \right), \label{eq:cvx_grad}
\end{align}
where $$\mathcal{A}^\star = \{ m \in \mathcal{T} \, \colon \, \phi_m( \theta^\star ) = \phi_{\mathcal{T}}( \theta^\star ) \}$$ is called the active index-set of the function $\phi_{\mathcal{T}}$ at $\theta^\star$ and the symbol $\text{co}\,C$ denotes the convex hull of the set $C$.

Identity~\eqref{eq:cvx_grad} thus says that the zero vector $0$ can be written as a convex combination of $| \mathcal{A}^\star |$ vectors, each vector pulled from a sub-differential~$\partial \phi_m( \theta^\star )$. Because the identity~\eqref{eq:cvx_grad} lives in~$\mathbf{R}^d$, Carathéodory's Theorem (Theorem 1.3.6 in~\cite{hiriart2004fundamentals}) asserts that there exists $\mathcal{A} \subset \mathcal{A}^\star$ such that \begin{align}
    0 \in \text{co} \left( \bigcup_{m \in \mathcal{A}} \partial \phi_m( \theta^\star ) \right), \label{eq:cvx_grad2}
\end{align}
and $| \mathcal{A} |  \leq d+1$.

Next, we consider two cases: 
\begin{itemize}
\item Case 1: $| \mathcal{A} |  = d+1$.  We define $\mathcal{S} = \mathcal{A}$;
\item Case 2: $| \mathcal{A} |  < d+1$. 
Note that $\mathcal{A} \subset \mathcal{T}$ and $| \mathcal{T} | = M-O$; thus, due to the assumption $d + 1 < M - O$, we can enlarge $\mathcal{A}$ with enough indexes from ${\mathcal T} \backslash \mathcal{A}$ so as to attain a set $\mathcal{S}$ of size $d+1$. 
\end{itemize}
Regardless of which case holds, we can thus generate $\mathcal{S}$ of size $d+1$ that satisfies \begin{align} 
\mathcal{A} \subset \mathcal{S} \subset {\mathcal T}.
\label{eq:subsup} 
\end{align}

Now, let the active index-set of the function
$$\phi_{\mathcal S}(\theta) = \max\{ \phi_m(\theta) ; \, m \in \mathcal{S} \}, \quad \text{for all }\theta,$$
at $\theta^\star$, be denoted by $\mathcal{B}^\star$, that is, $$\mathcal{B}^\star = \{ m \in \mathcal{S} \, \colon \, \phi_m(\theta^\star) = \phi_{\mathcal{S}}( \theta^\star ) \}.$$
It follows from~\eqref{eq:subsup} that $\mathcal{A} \subset \mathcal{B}^\star$, which, in view of~\eqref{eq:cvx_grad2}, implies
\begin{align}
    0 \in \text{co} \left( \bigcup_{m \in \mathcal{B}^\star} \partial \phi_m( \theta^\star ) \right). \label{eq:cvx_grad3}
\end{align}
By Corollary 4.3.2 in~~\cite{hiriart2004fundamentals}, 
identity~\eqref{eq:cvx_grad3} implies $0 \in \partial \phi_{\mathcal{S}}( \theta^\star )$, and, finally, Theorem 2.2.1 in~\cite{hiriart2004fundamentals} shows that~\eqref{eq:goal} holds. 

\section{Applications}
\label{sec:stromberg}
We discuss two applications that show the usefulness of Theorem~2.

\paragraph{Stromberg's Method}
In~\cite{stromberg1991computing}, Stromberg addresses a least-squares problem with outliers. The problem is formulated as~\eqref{eqn:percentile_min}, with convex residuals given by $f_m(\theta)=| y_m - x_m^T\,\theta |$. Furthermore, the condition~$d + 1 \leq M - O$ is assumed.

The method proposed in~\cite{stromberg1991computing} can be summarized as follows:
\begin{itemize}
\item Enumerate all subsets $\mathcal{S}$ of $\mathcal{M}$ of size $d+1$;
\item For each such subset $\mathcal{S}$, solve~\eqref{eqn:pfit}. Under the so-called Haar's condition invoked in~\cite{stromberg1991computing}, problem~\eqref{eqn:pfit} has a unique solution, say, $\text{Argmin}( \phi_{\mathcal{S}} ) = \{ \theta_{\mathcal{S}}^\star \}$;
\item Plug each $\theta_{\mathcal{S}}^\star$ into the objective of function interest~\eqref{eqn:percentile_min} and let the returned $\theta^\star$ be the one with the lowest value.
\end{itemize}

Stromberg~\cite{stromberg1991computing} showed that such $\theta^\star$ indeed solves~\eqref{eqn:percentile_min}. This was achieved by using the unique form of the residuals, specifically $f_m(\theta) = | y_m - x_m^T \theta |$, to characterize the solution set of Chebychev problems -- see~\cite{rivlin1963overdetermined,cheney1966introduction}. Theorem 2, on the other hand, is valid for arbitrary convex residuals $f_m$. Theorem 2 is able to capture the core insight of Stromberg's method by exploring the sub-differential characterization for the maximum of convex functions -- see equation~\eqref{eq:cvx_grad} in the proof of Theorem 2.

\paragraph{Robust centroid}
To illustrate the usefulness of Theorem 2 for more general convex residuals, we consider the problem of computing a centroid of $M$ points, of which $O$ are outliers. 
Letting $x_m \in \mathbf{R}^d$, for $1 \leq m \leq M$, denote the given points, we can phrase this problem as in~\eqref{eqn:percentile_min}, by considering convex residuals of the form $f_m(\theta) = \left\| x_m - \theta \right\|_2^2$; here, $\left\| \cdot \right\|_2$ denotes the $\ell_2$ (Euclidean) norm. In this setting, solving~\eqref{eqn:percentile_min} can be interpreted as computing the mean of a population, robustly against $O$ outliers~\cite{rousseeuw1984least}. 

Theorem 2 shows that~\eqref{eqn:percentile_min} can be solved by first solving $\mathcal{S}$-fits as in~\eqref{eqn:pfit}, thus accessing $\text{Argmin}( \phi_{\mathcal{S}} )$ for all $\mathcal{S}$ of size $d+1$, and then searching among those sets for the point that yields the smallest value of the function of interest~\eqref{eqn:percentile_min}.  For the problem of robust centroid at hand, it can be shown that each $\mathcal{S}$-fit has a unique solution, say, 
$\text{Argmin}( \phi_{\mathcal{S}} ) = \left\{ \theta_{\mathcal{S}}^\star \right\}$, thanks to the strict convexity of the residuals. 
Furthermore, because we take $d = 2$ for ease of visualization, each $\theta_{\mathcal{S}}^\star$ can be easily computed resorting to techniques in~\cite{kallberg2019minimum}. 

\begin{figure}[h!]
	\centering
	\begin{subfigure}{0.5\textwidth}
		\centering
		\includegraphics[width=5.5cm]{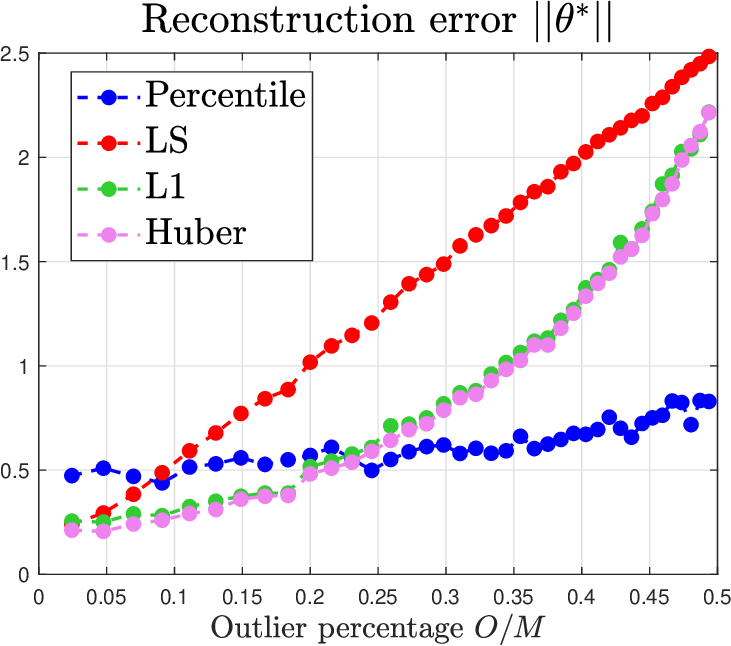}
		\caption{ Average error over $100$ Monte Carlo trials.  }
		\vspace{0.4cm}
		\label{fig_a}
	\end{subfigure}
	\begin{subfigure}{0.5\textwidth}
		\centering
		\includegraphics[width=4.5cm]{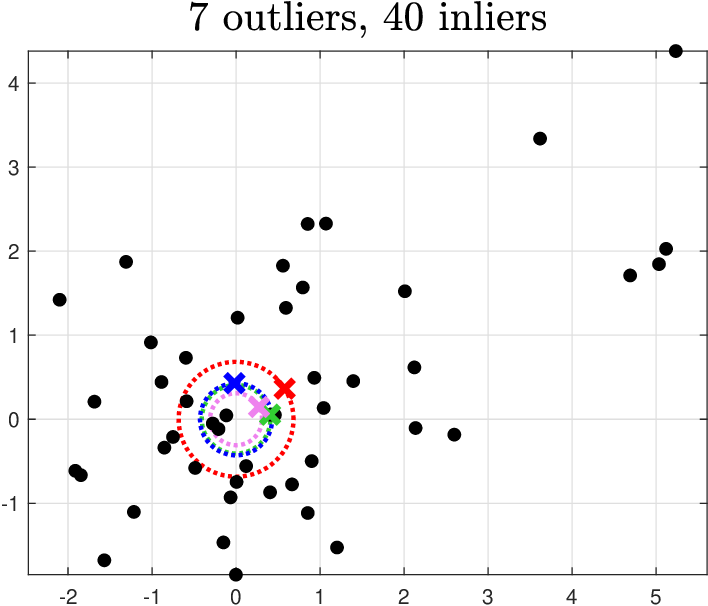}
		\caption{ A typical realization with $15\%$ outlier ratio.  }
		\vspace{0.4cm}
		\label{fig_b}
	\end{subfigure}  
	\begin{subfigure}{0.5\textwidth}
		\centering
		\includegraphics[width=4.5cm]{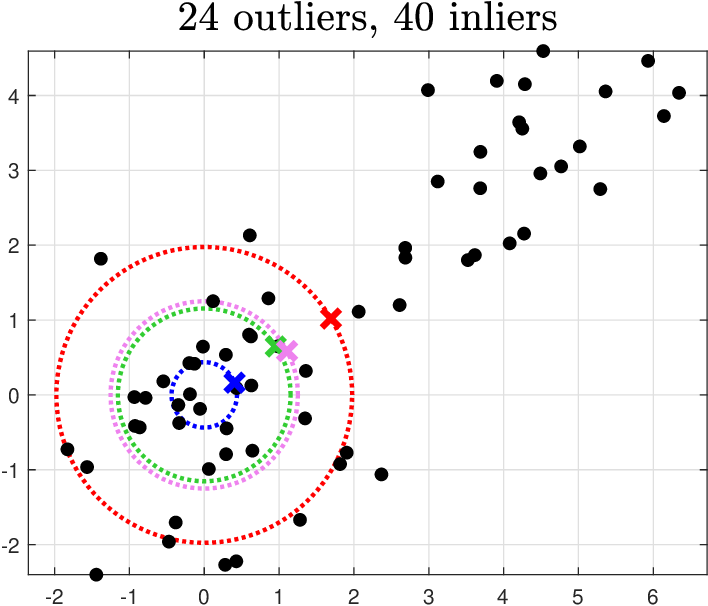}
		\caption{ A typical realization with $38\%$ outlier ratio.  }
		\label{fig_c}
	\end{subfigure}
	\caption{Results for the robust centroid example: panel (a) shows that, when the fraction of outliers exceed $23\%$, only the percentile method based on Theorem 2 prevails.}
	\label{fig:numerical_figure}
\end{figure}

For our numerical experiments, we start by sampling $M - O = 40$ (inlier) points from a normal distribution $\mathcal{N}(0,I)$. Then, we add outliers by sampling $O$ points from a shifted normal distribution with higher variance $1.2\,\mathcal{N}(b,I)$, where the bias vector $b = (4, 3)$.
Figures~\ref{fig:numerical_figure} (b) and (c) plots realizations of this setup for an increasing number of outliers~$O$.

On this dataset, we compare the method enabled by Theorem 2 (delineated above) with the least squares solution~\eqref{eqn:LS_formulation} and two classical methods~\cite{huber2004robust,candes} from robust estimation:
\begin{align*}
		&\min_{\theta } \sum_{m=1}^M || x_m-\theta||_1 \enspace \enspace \enspace \enspace \enspace {(\text{L}1)}, \\
		&\min_{\theta } \sum_{m=1}^M h_R( \left\|x_m-\theta\right\|_2 ) \enspace \enspace {(\text{Huber})}, 
\end{align*}
where $\left\| \cdot \right\|_1$ denotes the $\ell_1$ norm and $h_R$ the Huber function with threshold $R$ (we set $R=1.34$ as suggested in~\cite{huber2004robust}, since the standard deviation of inliers is unitary).
~\\
In view of the way the data set was generated, we wish the methods to return the zero vector $(0,0)$, as this is the (theoretical) mean of the population of inliers. Thus, we compare the four methods on the basis of the Euclidean norm of their returned parameter $\theta^*$.

Figure~\ref{fig:numerical_figure} gives the results of the comparison, as the number of outliers increases. Figure~\ref{fig:numerical_figure} (a) shows that the percentile method (in blue) is the only method that endures a moderate-to-high number of outliers, say when $O/M > 23\%$. 
These findings confirm that typical robust alternatives cannot withstand considerable amounts of outliers, which highlights the importance of Theorems~1 and~2 for these challenging setups.





\bibliographystyle{IEEEtran}

\bibliography{IEEEabrv,02.biblio}


\end{document}